# Electrically tunable giant Nernst effect in two-dimensional van der Waals heterostructures

Gabriele Pasquale[1,2], Zhe Sun[1,2], Kenji Watanabe[3], Takashi Taniguchi[4], Andras Kis[1,2*]

[1]*Institute of Electrical and Microengineering, École Polytechnique Fédérale de Lausanne (EPFL), CH-1015 Lausanne, Switzerland*

[2]*Institute of Materials Science and Engineering, École Polytechnique Fédérale de Lausanne (EPFL), CH-1015 Lausanne, Switzerland*

[3]*Research Center for Electronic and Optical Materials, National Institute for Materials Science, 1-1 Namiki, Tsukuba 305-0044, Japan*

[4]*Research Center for Materials Nanoarchitectonics, National Institute for Materials Science, 1-1 Namiki, Tsukuba 305-0044, Japan*

*Correspondence should be addressed to: Andras Kis, andras.kis@epfl.ch

**ABSTRACT**

**The Nernst effect, a transverse thermoelectric phenomenon, has attracted significant attention for its potential in energy conversion, thermoelectrics, and spintronics. However, achieving high performance and versatility at low temperatures remains elusive. Here, we demonstrate a large and electrically tunable Nernst effect by combining graphene's electrical properties with indium selenide's semiconducting nature in a field-effect geometry. Our results establish a novel platform for exploring and manipulating this thermoelectric effect, showcasing the first electrical tunability with an on/off ratio of $10^3$. Moreover, photocurrent measurements reveal a stronger photo-Nernst signal in the Gr/InSe heterostructure compared to individual components. Remarkably, we observe a record-high Nernst coefficient of $66.4\ \mu V\ K^{-1} T^{-1}$ at ultra-low temperatures and low magnetic fields, paving the way toward applications in quantum information and low-temperature emergent phenomena.**

## INTRODUCTION

The investigation of thermoelectricity traces its origins back to the mid-nineteenth century when Lord Kelvin embarked on a quest to comprehend it as a quasi-thermodynamic phenomenon. A significant milestone in this journey occurred in 1931 with the formulation of reciprocal relations by Onsager[1]. Such relations established crucial connections, including the Kelvin relation between Seebeck and Peltier coefficients and the Bridgman relation linking the Nernst and Ettingshausen effects[2]. Practical applications, however, have been limited to date. Nevertheless, recent technological advancements and promising applications in energy conversion, thermoelectrics, and spintronics have renewed interest in thermoelectric phenomena[3–7]. One such effect is the Nernst-Ettingshausen effect, which manifests itself as a transverse electric field, known as the Nernst voltage, generated by the Lorentz force acting on charge carriers in the presence of a temperature gradient and a magnetic field. Among the recently investigated materials, topological semimetals show promise for efficient thermoelectric cooling via the Nernst-Ettingshausen effect[3,8]. Such materials are characterized by zero or slight band overlap, and high carrier mobilities that are beneficial to enhancing thermoelectric effects. On the other hand, difficulties in measuring the transverse thermoelectric effects have slowed down the progress compared to its longitudinal counterpart[9], despite extensive work on reaching high thermoelectric figures of merit[3,5,6,10–12].

Achieving a sizable and tunable Nernst effect at ultra-low temperatures remains an ongoing challenge[13,14], especially in the sub-Kelvin regime[10], where it is gaining increasing momentum due to potential applications in quantum technologies. Within the context of qubit circuits, where precise thermal control is paramount[15,16], the ability to convert localized heat sources, both internal and external to the circuit, into controllable electric signals emerges as a pivotal asset[15,17]. Hence, this capability has the potential to contribute to the fine-tuning of quantum systems, although the full extent of its impact and the complexities involved in manipulating quantum states remain subjects of ongoing research.

To date, implementing such low-temperature thermal management methods remains challenging due to significant magnetic field constraints, thus requiring advancements in materials and techniques. In fact, conventional materials exhibit limited Nernst responses at ultra-low temperatures[12,18], motivating the search for novel materials with large and tunable Nernst effects in such conditions[18].

Here, we show a large and tunable Nernst effect by combining graphene with the metal monochalcogenide InSe assembled in a field-effect geometry. After the observation of the photo-induced Nernst effect in graphene[19], further studies have provided insights into the nature of the effect and its applications[20]. Among several candidates, InSe is chosen due to a combination of desirable properties, such as high electron mobility[21], low resistivity, and because of its peculiar band topology, which is predicted to give rise to enhanced thermoelectric phenomena[22–31]. By taking advantage of the exceptional electrical conductivity of graphene and the intriguing semiconducting properties of indium selenide (InSe), we demonstrate the first electrically tunable Nernst effect with an on/off ratio of $10^3$ in a field effect structure. This signal arises when the sample is subject to a magnetic field and a temperature gradient generated by laser illumination. Furthermore, by measuring the signal in a different geometry, using graphene as electrodes and InSe as a channel material, we obtain a Nernst coefficient of $66.4\ \mu V K^{-1} T^{-1}$ which represents the highest value observed at ultra-low temperatures and low magnetic fields.

These findings position our device as a promising candidate for enabling highly efficient Nernst operation in low-temperature conditions[32–34]. Thus, by harnessing the unique characteristics of graphene and InSe, our study unveils a novel platform for exploring and manipulating the Nernst effect, potentially enabling transformative applications in the realm of low-temperature thermoelectric and spintronic devices.

## RESULTS

**Device structure and Nernst effect in Gr/few-layer InSe heterostructure**

To ensure the high quality and reliability of our devices, we employ a fabrication process that involves encapsulating γ-phase InSe flakes within hexagonal boron nitride (hBN) and using graphene or few-layer graphite (FLG) flakes as electrodes[35,36]. The device architecture is designed to incorporate a graphene electrode spanning the entire length of the InSe flake, enabling measurements of graphene properties and their response to the proximity of InSe[37]. Additionally, a FLG bottom gate is utilized to modulate the carrier density in the semiconductor. The schematic of the Gr/InSe heterostructure is illustrated in Figure 1a.

We present the longitudinal transport characteristics of a representative device in Figure 1b, demonstrating clear ambipolar transport in few-layer InSe[31]. We perform thermoelectric power measurements at 100mK in a dilution refrigerator by performing scanning photocurrent (and photovoltage) measurements, both in DC and in AC using a lock-in amplifier (see Methods and Supplementary Note 6). Figure 1c depicts a false-color image of a 3-layer InSe/Gr device, with a superimposed spatial map illustrating the photo-Nernst effect (PNE) signal within the field of view. The PNE effect is characterized by the emergence of a transverse current (voltage) upon laser illumination at the edges of the channel[19]. Such illumination induces an uncompensated thermal gradient that drives the current along the channel when the sample is subjected to a magnetic field[19,20]. Importantly, this signal changes sign at opposite magnetic field polarities and remains consistent across the channel's length as the laser is scanned, exhibiting a uniform profile that changes sign at opposite edges of the flake, in accordance with the principles outlined in the Shockley-Ramo theorem[38].

The photo-Nernst effect (PNE) current can be described by the equation: $I_{ph} = \beta N B \rho_{xx}^{-1} \cdot \Delta T_{av}$, where N represents the Nernst coefficient, B the applied magnetic field perpendicular to the channel, $\rho_{xx}^{-1}$ the inverse of the longitudinal resistivity, and $\Delta T_{av}$ the average temperature difference induced across the edges by means of laser illumination, while $\beta < 1$ accounts for the geometric factor that incorporates contact resistance[19]. Throughout our work, we conduct measurements of both PNE current and voltage, with a detailed derivation and formalism presented in Supplementary Note 3. To confirm the nature of the effect, we demonstrate the linear and antisymmetric behavior of the PNE effect with respect to the applied magnetic field, and the linearity of the effect by increasing the laser power[19], as depicted in Figure 1d. Additional characterization and measurements can be found in Supplementary Note 3. Notably, the PNE effect can be measured both on bare graphene and within the heterostructure region, emphasizing that the presence of InSe does not quench the signal. Importantly, all measurements are conducted with a laser wavelength of $\lambda = 532\ nm$, excitation power of 50 μW, and under a magnetic field of 1 Tesla unless stated otherwise. At such laser powers, the background temperature of the sample stabilizes around 100 mK.

To evaluate the Nernst response of the heterostructure, we compare the PNE measured when shining the laser light on the bare graphene, with that obtained from the Gr/InSe heterostructure as a function of gate voltage, as shown in Figure 2a. The graphene signal aligns with previously reported findings[19,20], exhibiting a peak feature at the Dirac point and decaying branches for both voltage polarities. In contrast, the signal within the Gr/InSe structure undergoes a dramatic change when varying the gate voltage from positive to negative values. To understand this result, we perform an in-depth analysis of the effect on the heterostructure. We record the PNE signal as a function of the magnetic field and gate voltage, observing the expected linear and antisymmetric behavior as measured previously for bare graphene (Figure 1d). However, the slope of this effect shows significant changes upon modulation of the gate voltage. In particular, at positive gate voltages where InSe exhibits high conductivity, the PNE effect is strongly suppressed. Conversely, at increasingly negative gate voltages, the slope increased well beyond the value observed at $V_g = 0$ V. By plotting the PNE intensity at 1T of the Gr/InSe heterostructure against gate voltage on a logarithmic scale, a clear modulation of the effect that can be switched on and off is revealed, with a ratio of $\sim 10^3$ (Figure 2b). Since no quenching of the effect is observed on the bare graphene, as previously investigated[19,20], we attribute this effect to the presence of the InSe flake, as further elaborated below.

We can compute the Seebeck coefficient of InSe using the traditional Mott relation[39]: $S_{InSe} = \frac{\pi^2 k_b^2 T}{3e} \frac{1}{G} \frac{dG}{dV_g} \frac{dV_g}{dE_F}$, where $k_b$ is the Boltzmann constant, $T$ is the temperature, $e$ the electron charge, $G$ the conductivity, $V_g$ the gate voltage and $E_F$ the Fermi energy. The factor $\frac{dV_g}{dE_F}$ was calculated based on our previous work[22] (refer to Supplementary Note 5 for detailed information). We acknowledge that while the Mott formula may not yield results in perfect agreement with the experimental ones as reported previously[28], the observed discrepancies typically remain within a factor of 2. Importantly, these deviations do not alter the core findings and conclusions of our work, namely an electrically tunable Nernst device operating at low magnetic fields and millikelvin temperatures, as discussed further in Supplementary Note 3.

By keeping into account the gate dependence of the Seebeck effect, we can derive the values for the Seebeck coefficient of InSe which are shown for room temperature (yellow) and for 100mK (Red) in Figure 2c (Supplementary Note 3). The room temperature values of the Seebeck coefficient of few-layer InSe are comparable to the best values reported for monolayer semiconducting materials[40,41]. Further, they exhibit a higher tunability ranging from a maximum value of $-3.7 \cdot 10^5 \mu V/K$ to a minimum value of $-4.1 \cdot 10^{-2} \mu V/K$. Thus, a difference of seven orders of magnitude is present, compared to three orders of magnitude as shown for e.g. monolayer $MoS_2$[41]. We note that our devices are in the few-layer thickness regime (3-5L), and we anticipate that devices with fewer layers could exhibit even more pronounced effects[30]. At low temperatures, the tunability remains significant, although the values are noticeably reduced compared to room temperature.

The high Seebeck coefficient of few-layer InSe relative to graphene gives rise to a substantial thermoelectric voltage at the interface upon laser illumination that acts as an additional electric bias, resulting from the thermoelectric signal denoted by $V_{th} = \Delta T \cdot (S_{InSe} - S_{Gr})$ [41,42]. Hence, we can qualitatively elucidate the quenching of the PNE signal at the interface for positive gate voltages, as the Seebeck coefficient of InSe becomes greatly reduced and comparable to that of graphene, leading to the suppression of the thermoelectric voltage caused by the disparity in Seebeck coefficients[43]. Conversely, the highly resistive state of InSe within the bandgap and its hole conduction state facilitates a substantial Seebeck coefficient, thereby preserving the thermal gradient at the interface. Furthermore, the presence of defects and impurities in the InSe layer can also alter the electronic properties of graphene. In such a process, if the charge carriers photo-generated in the graphene layer scatter off the defects and impurities in the InSe layer, it would effectively increase the local temperature when the Fermi level lies below the energy of the defect states in InSe (see Supplementary Note 2 for further details).

From the PNE equations (Equations S9 and S10 in Supplementary Note 3), we can derive the lower bound of the Nernst coefficient for InSe. The values of $\beta N$ obtained for the 3L InSe/Gr heterostructure are presented in Figure 2d as a function of gate voltage. The Nernst coefficient can be effectively tuned by manipulating the carrier density, reaching a maximum value of $17.5\ \mu V/K$ at negative voltages and becoming negligible at positive gate voltages. The error bar in the plot represents the effect of the error of the thermal gradient propagated to the Nernst coefficient. Remarkably, this wide range of tunability for the Nernst coefficient can enhance the Nernst effect of graphene to compete with the best materials available at comparable temperatures, namely Bi and Bi-Sb alloys[14,44]. Hence, it holds great promise for applications of the Nernst effect in miniaturized devices, enabling the active switching of thermopower efficiency through an applied electric field.

**Giant Nernst response in field effect geometry**

During our investigation, we noted the emergence of a Nernst signal when employing an alternative measurement geometry, as illustrated in Figure 3a. This configuration involves the utilization of two graphene flakes as electrodes positioned on the InSe semiconducting channel, forming a field effect structure. The system is illuminated with laser light, and a laser reflectance map of the device is shown in Figure 3b, where the two positions labeled on the map represent the measurements sites on bare graphene (1) and the graphene/InSe heterostructure region (2). To quantify the effect in this geometry, we compare the signals obtained by illuminating the graphene alone (site 1) and the heterostructure (site 2) while varying the magnetic field (Figure 3c). Notably, the signal recorded on the graphene electrode is comparable to that of the proximitized graphene discussed earlier (Figure 2a). On the other hand, the signal recorded on the heterostructure shows a clear deviation from linearity when the magnetic field goes above $\pm\, 0.5\, T$. To compare the effects, we restrict ourselves to the linear region, and we compute the slope of the photocurrent. Such a slope serves as a parameter for assessing the efficiency of the two effects, with a higher slope indicating a stronger effect. Remarkably, when the laser is directed onto the heterostructure, a significantly larger signal is observed with respect to the first site, and the ratio of the two slopes is approximately 42. The result is reproducible over several locations on the sample. This suggests that the InSe channel facilitates a more efficient effect overall, yielding a calculated lower bound of the Nernst coefficient of $66.4\, \mu V K^{-1} T^{-1}$ at 1T.

To gain further insight, we record the scanning photocurrent map in the absence of bias and gate voltage. In particular, when the magnetic field is applied, we observe a strong signal arising on one of the graphene electrodes, as depicted in scanning photocurrent maps presented in Figures 3d, 3e, and 3f, corresponding to -1T, 0T, and 1T magnetic fields, respectively. While a weak signal is discernible on the graphene electrode itself, the device geometry constrains the photocurrent to flow through the InSe channel, making the signal more challenging to detect. The observed signal arises upon application of a magnetic field, and both the trend and the sign of the measured photocurrent follow the photo-Nernst effect. One possible origin of the effect can be attributed to the known favorable interplay between the low Fermi energy of the system, a lower thermal conductivity with respect to graphene[28,29,45,46], and its high electron mobility[6,23,25]. However, the precise mechanism behind this phenomenon requires further investigation. To gain a comprehensive understanding of the microscopic origins of this enhancement, rigorous theoretical investigations are also encouraged.

**Benchmarking of thermoelectric properties**

One crucial parameter that allows for the observation of the Nernst-like effect in the geometry shown in Figure 3a is the high electron mobility of our InSe channel. To evaluate the performance of our field-effect devices in comparison to the existing literature, we extract the room temperature field-effect mobility and current on/off ratio for our InSe-based devices (Figure 4a). Our devices exhibit superior performance in both the current on/off ratio and field-effect mobility compared to previous reports. Notably, we achieve a record on/off ratio of approximately $10^7$ and maximum field-effect mobility of approximately $150\ cm^2V^{-1}s^{-1}$, surpassing the best values reported so far of approximately $10^5$ and $10\ cm^2V^{-1}s^{-1}$, respectively (see Figure 4a)[47]. This performance enhancement results from improved InSe material quality (HQ Graphene, see Methods) and device fabrication techniques. The high electron mobility in our devices enables the observation of the photo-Nernst effect for the first time through a layered semiconductor channel, facilitated by the low resistivity of the few-layer InSe, which is comparable to that of graphene without an applied gate voltage[48]. In particular, we plot the Nernst coefficients in our devices and we compare them with the existing values present in the literature at their respective temperature, as depicted in Figure 4b. Our results show a wide electrical tunability of the Nernst effect across a wide range of values, reaching the best device performances reported so far. Moreover, the value of the Nernst coefficient obtained in the heterostructure region when measuring in the geometry shown in Figure 3a as detailed above represents, to the best of our knowledge, the highest value observed for modest magnetic fields at temperatures below 200 mK. Thus, it establishes a new benchmark for the lowest operational temperature ever utilized successfully in Nernst measurements.

**CONCLUSION**

In summary, our study presents the demonstration of a micrometer-sized thermoelectric device harnessing the photo-induced Nernst effect, displaying exceptional performance even at ultra-low temperatures of 100 mK, which were previously unattainable. By harnessing the unique properties of the Gr/InSe heterostructure, we achieve a Nernst coefficient comparable to bismuth at ~200 mK and 1T, while benefiting from the added advantage of tunability through carrier concentration modulation. Thus, leveraging the two-dimensional nature of our device, we achieve the first low-temperature, high-performance, tunable thermoelectric Nernst effect in a field-effect device. We note that the main advantage of having electrical tunability relies in the control and versatility it offers in practical applications. Since it can be achieved through standard electronic components, electrical control is straightforward to integrate into electronic devices and systems, making it compatible with existing technology.

In the traditional field effect geometry (Figure 3a), we observe an unprecedented record-high Nernst coefficient of $66.4\ \mu V K^{-1} T^{-1}$ at 100 mK and 1T, which is comparable to the Nernst coefficient values of materials currently employed at room temperature in commercial devices[49]. These findings not only establish Gr/InSe heterostructures as promising candidates for ultra-low temperature operation and the investigation of emergent physics but also emphasize the significance of precise control and conversion of heat into electrical signals in such systems. In particular, the ability to convert localized heat sources into controllable electric signals could be implemented in qubit circuits as a thermal management technique, which is an active area of research efforts[15,17].

By effectively bridging the gap between fundamental research and practical applications, our work provides a solid foundation for transformative advancements in quantum technologies, emergent phenomena, and thermoelectric engineering[32–34].

**ACKNOWLEDGEMENTS**

We acknowledge helpful discussions with Dr. Guilherme Migliato Marega, Edoardo Lopriore, and Fedele Tagarelli. We acknowledge the support in microfabrication and e-beam lithography from EPFL Centre of MicroNanotechnology (CMI) and thank Z. Benes (CMI) for help with electron-beam lithography. This work was financially supported by the Swiss National Science Foundation (grant nos. 175822 and 164015), the European Union's Horizon 2020 research and innovation program under grant agreement 881603 (Graphene Flagship Core 3). K.W. and T.T. acknowledge support from the JSPS KAKENHI (Grant Numbers 21H05233 and 23H02052) and World Premier International Research Center Initiative (WPI), MEXT, Japan.

## METHODS

### Device fabrication

The heterostructures utilized in this study were fabricated using the conventional dry transfer method. Initially, the hBN and graphene/few-layer graphene (NGS) building blocks were obtained by mechanical exfoliation and deposition onto silicon oxide. Subsequently, all the components were assembled by starting from the uppermost hBN layer, which was lifted using a polycarbonate (PC) membrane on PDMS and carefully placed on top of a few-layer graphene (FLG) bottom gate. The few-layer InSe (HQ Graphene) flakes were exfoliated onto PDMS (Gelpak) and distinguished based on their optical contrast. To prevent material degradation and contamination, all of these procedures were carried out within an argon-filled glovebox (Inert).

Once the sample was fully encapsulated, it underwent annealing at 340 ºC in a high vacuum with a pressure of $10^{-6}$ mbar for a duration of 6 hours. Lastly, electrical contacts were fabricated by employing e-beam lithography and depositing metal (Ti/Au) through evaporation with thicknesses of 2 nm and 100 nm, respectively.

**Optical and electrical measurements**

All measurements shown in this work were carried out under vacuum at 100mK unless specified otherwise. Scanning photocurrent and laser reflectance measurements were performed by focusing a laser on a spot of about 1 μm diameter on the sample. Multiple laser sources have been used for this purpose, and consistent results have been obtained for all the sources: a narrow-linewidth tunable continuous wave laser (MSquared), CW laser diodes (Thorlabs) with wavelength centered at 780nm, 904nm, 648nm, for Nernst measurements, photocurrent measurements, and thermal signal generation. In particular, for all the measurements reported in the main text a laser wavelength of 532nm of a CW laser is used. The incident power was varied from 1 μW to 300 μW for power dependence measurements and kept at 50 μW for the PNE measurements shown in the main text unless otherwise specified. Transport measurements were carried out at room temperature, and 80 mK with a Keithley 2636 sourcemeter. The 80 mK temperature was achieved inside a dilution fridge from Oxford Instruments, with a custom-made window and mirrors that allow us to perform optical and optoelectronic measurements. The background temperature is affected by the laser power, with a stable temperature of 100 mK when employing 50 μW. The electrical signal was detected both by DC and AC photocurrent (and photovoltage) with a Stanford Research lock-in amplifier SR830, driven by a frequency of 727Hz given by a synchronized laser chopper.

## AUTHOR CONTRIBUTIONS

A.K. initiated and supervised the project. G.P. fabricated the devices. K.W. and T.T. grew the h-BN crystals. G.P. performed the optical and electrical measurements, assisted by Z.S. G.P. analyzed the data with input from A.K. G.P. wrote the manuscript, with inputs from all the authors.

## COMPETING INTERESTS

The authors declare no competing financial or non-financial interests.

## DATA AVAILABILITY

The data that support the findings of this study are available from the corresponding author on reasonable request.

**FIGURES**

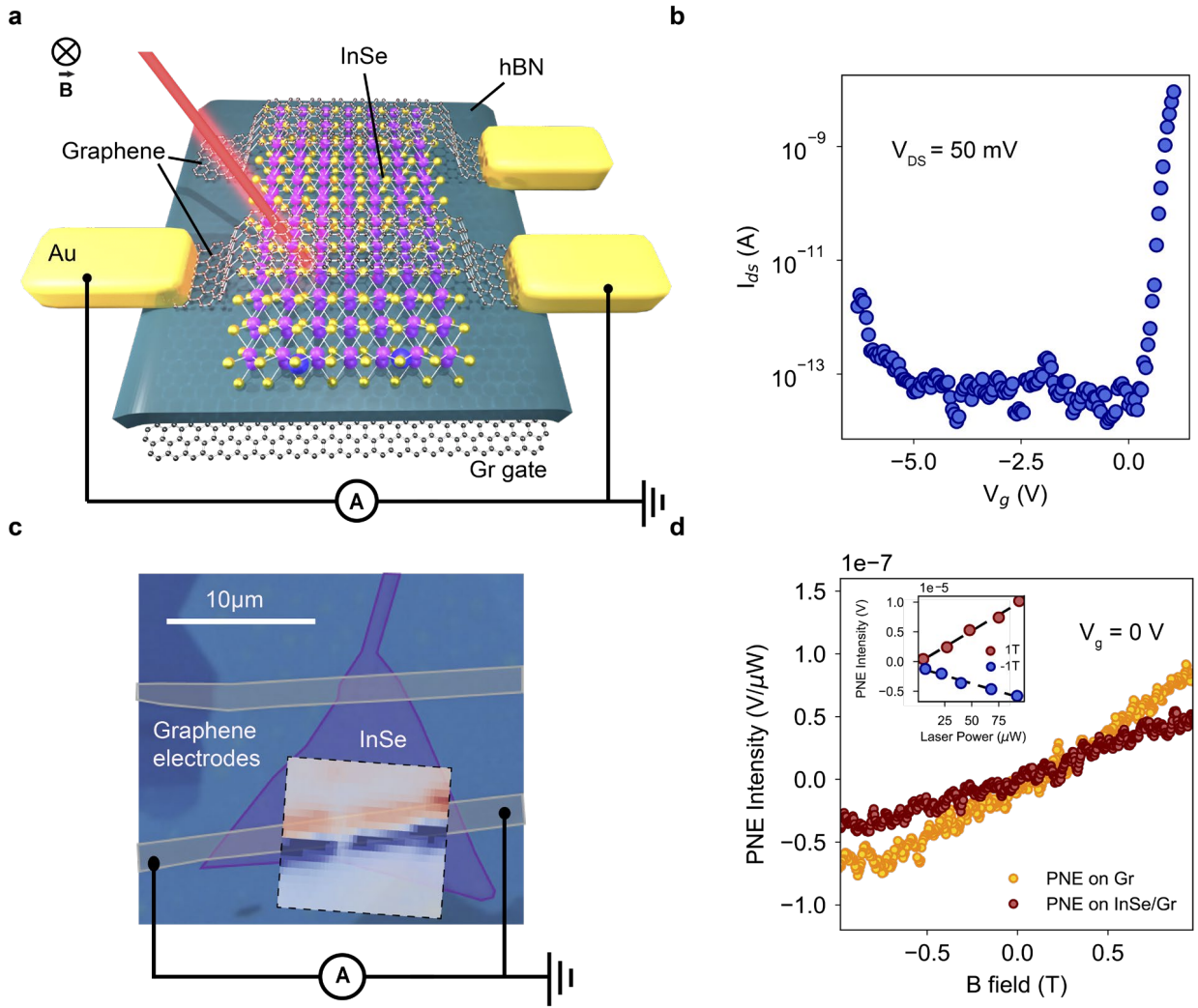

**Figure 1. Device schematics and basic characterization**. (a) Device schematic representing a fully encapsulated few-layer InSe channel, with graphene electrodes. One of the electrodes is contacted on both sides with gold contacts in order to perform measurements of graphene, affected by the InSe. The scanning photocurrent maps are performed at 100mK, with 50µW laser power and an out-of-plane magnetic field of 1T, unless specified otherwise. (b) Transfer characteristic of a typical few-layer InSe device, performed at low temperature under 50mV of bias. (c) Optical micrograph of a 3L InSe device with graphene electrodes. An exemplary measurement of the photo-Nernst effect at 1T measured on graphene is shown overlapped to the region of interest. (d) Linear magnetic field dependence of the PNE when the illumination is upon graphene alone (dark yellow) and InSe/Gr (dark red), respectively. The inset shows the power dependence of the PNE and sign change for an opposite magnetic field, confirming the origin of the effect as the photo-Nernst.

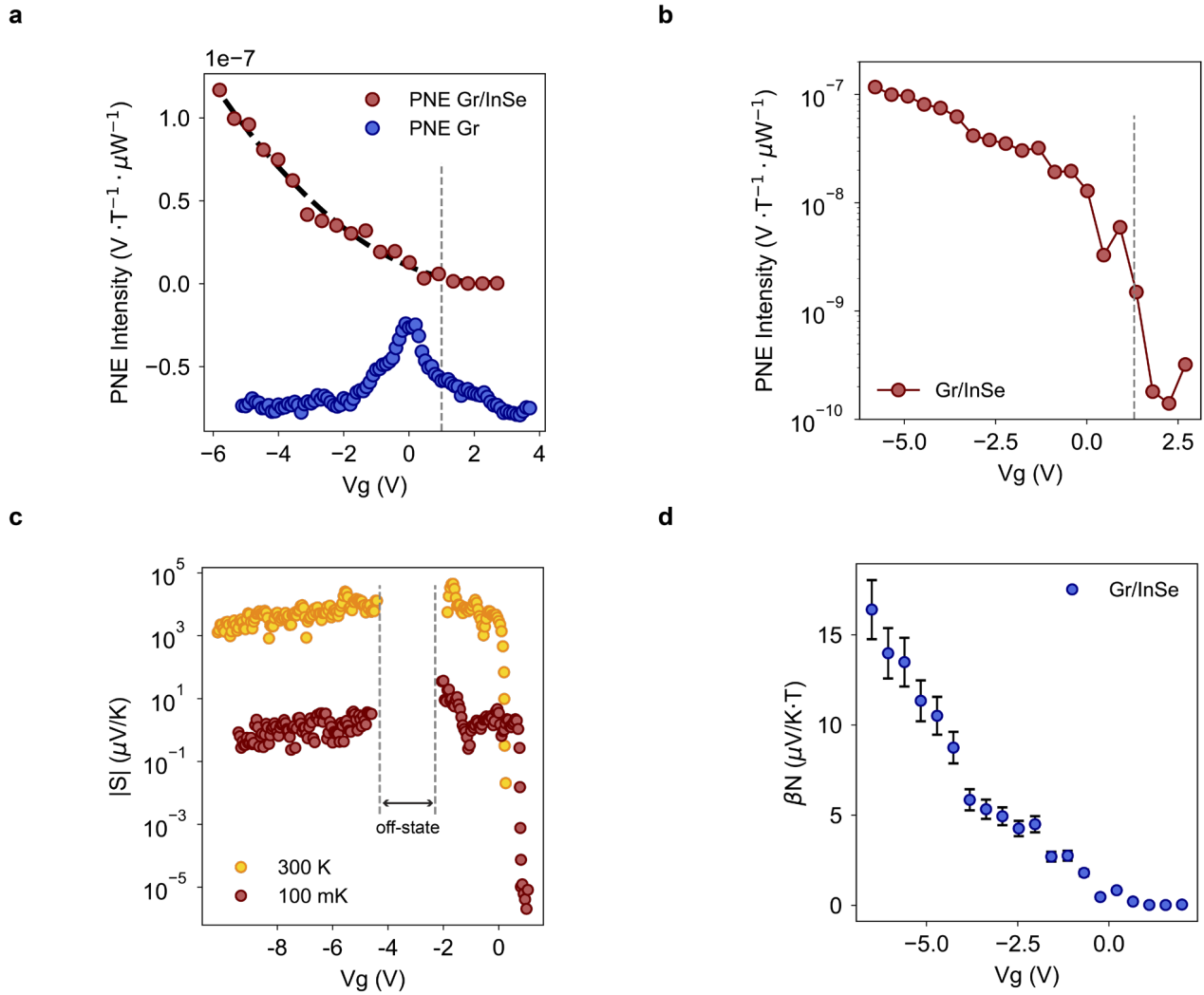

**Figure 2. Photo-Nernst effect and thermoelectric performance:** (a) PNE intensity normalized by laser power measured on the Gr/InSe heterostructure (dark red) as a function of gate voltage. The PNE of bare graphene is shown as a comparison (blue). (b) PNE intensity of the heterostructure plotted in logarithmic scale to highlight the sizeable on/off ratio. The effect can be switched on and off by changing the carrier density within our device. The gray dashed lines represent the onset of n-type conduction, analogously to (a). (c) Seebeck coefficient calculated through the Mott relations (see Supplementary Note X) as a function of gate voltage, shown for 300K (yellow) and 100mK (red). The vertical gray dashed lines represent the off-state of the device, where the Seebeck coefficient cannot be defined due to the high resistance of the device, which becomes comparable to the input impedance of the instrument[40]. (d) Gate tunable Nernst coefficient in the Gr/InSe heterostructure. The plot represents a lower bound for the real value, due to β. The error bar reflects the error present in the determination of the temperature, for further detail see Supplementary Note X.

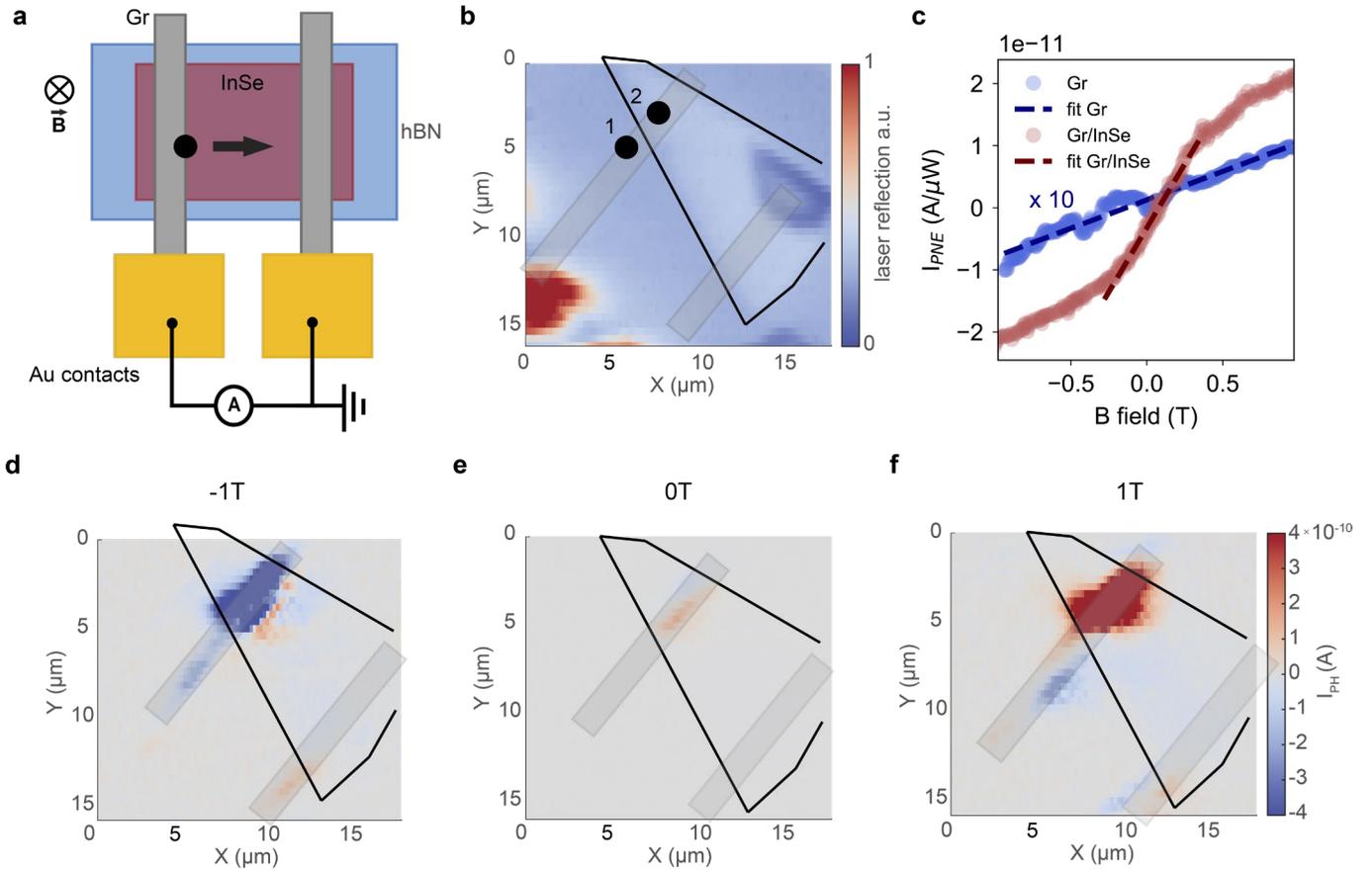

**Figure 3. Scanning photocurrent map of the PNE across an InSe channel:** (a) Device schematic showing the illumination of the Gr/InSe heterostructure and electrical detection across the InSe channel. Following this schematic, any measured current is forced to flow through the semiconductor. (b) Laser reflectance map of the region of interest measured simultaneously with the scanning photocurrent map. This measurement allows us to correlate the position of the laser with the signal observed. The positions chosen to record the PNE signal on graphene and on the InSe/graphene heterostructure are labeled as position 1 and 2, respectively (c) PNE signal recorded varying magnetic field and under 50μW of laser illumination and $V_g = 0\ V$ in position 1 and 2, shining light on the graphene electrode and on the heterostructure, respectively. The bare graphene signal is shown in blue, magnified by an order of magnitude to better highlight the difference in slope between the two curves. The measurements are performed without any applied bias as it would obscure the PNE, inducing other photocurrent mechanisms in the picture. (d) A scanning photocurrent map showing the measured photocurrent across the full device at -1T of applied out-of-plane electric field. (e) and (f) analogous scanning photocurrent map shown for 0T and 1T, respectively.

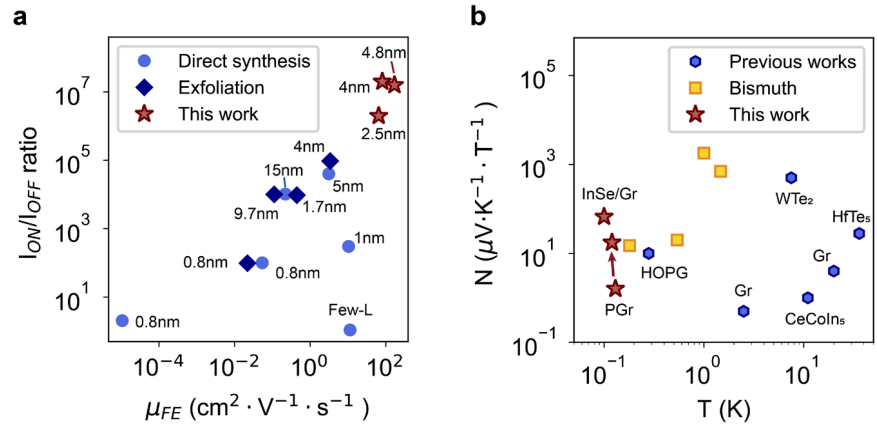

**Figure 4. Thermoelectric benchmarking**: (a) Benchmarking of the field-effect two-terminal mobility and on/off ratio of InSe-based devices, as recently reported in [47]. The values measured in this work are reported as red stars, showing a substantial improvement in both the on/off ratio and in the field effect mobility. (b) Low-temperature values of the Nernst coefficient for different material platforms[5,11,12,14,18,43,50–52]. In particular, bismuth is shown as yellow squares, since it possesses the best performance reported to date. The values measured in this work are shown as red stars. The red arrow indicates the tunability of the Nernst coefficient measured on graphene in proximity with InSe when changing the gate voltage, outperforming the bismuth counterpart. All values taken from the literature are displayed at 1T, to have a meaningful comparison, since most of these materials possess high Nernst values for different ranges of magnetic field.